# A Wideband Holographic Array with Azimuth and Elevation Beam Steering for 5G/6G Applications

Hazhir Mohammadi, Amir Saman Nooramin, Homayoon Oraizi, *Life Senior Member, IEEE*

*Abstract*— This paper presents the design and fabrication of leaky-wave holograms with anisotropic properties, enabling circular polarization for single-beam radiation in desired directions. Tailored for future terrestrial networks like 5G, 6G, and beyond, the proposed antenna achieves high-gain performance with beam steering in elevation and azimuth over a wide angular range. A planar reflector enhances directivity by suppressing backward modes, while a simple feeding network ensures optimal impedance matching. The antenna operates efficiently from 16 GHz to 20 GHz (22.22% bandwidth), delivering a gain of 24.5 dB, with compact dimensions of 15 cm × 17 cm. The structure includes a reflector, monopoles, a patterned patch, a Rogers 4003C substrate, a ground plane, and a coaxial feed. Simulations, experiments, and theoretical analyses confirm the design's efficacy, showcasing 2D scanning capabilities ideal for autonomous vehicle localization systems. It's simple, scalable design supports cost-effective mass production, offering significant potential for industrial applications.

*Index Terms*— Holographic Antenna, Azimuth beam steering, Elevation beam steering, Flat reflector antennas, High-gain antennas, Phased array antennas, Circular polarization, 5G and 6G antennas.

## I. INTRODUCTION

The emergence of 5G and the impending rollout of 6G technologies have significantly increased the demand for high-capacity, efficient communication systems. A critical challenge in these systems is the substantial propagation loss that increases with frequency. To mitigate this, antennas must be highly directive and capable of dynamically rotating and steering beams. As a result, holographic antennas have attracted attention for their potential in high-gain applications [1]. These antennas enable precise control over electromagnetic wave propagation, utilizing subwavelength structures to create compact, efficient designs for modern wireless communication systems. By using surface-impedance modulation to convert surface waves into space waves, holographic antennas allow the realization of miniaturized, functional systems. Earlier research focused on isotropic metasurfaces, but recent advancements have shifted towards anisotropic holographic antennas.

H. Mohammadi, A. Nooramin and H. Oraizi are with the Department of Electrical Engineering, Iran University of Science and Technology, Tehran 13114-16846, Iran. (e-mail: h_oraizi@iust.ac.ir)

These structures offer enhanced control over wave manipulation, including polarization, beam shaping, and multi-dimensional scanning, opening new avenues for antenna design. Anisotropic holographic antennas allow for direction-dependent modulation of surface impedance, offering improved control over polarization, nonreciprocal radiation, and the ability to manipulate beam direction in multiple dimensions. For example, [2] demonstrated the use of modulated surface impedance in spiral leaky-wave antennas, achieving superior circular polarization and directivity compared to traditional leaky-wave antennas. Similarly, [3] introduced metasurface-based planar antennas with tunable radiation patterns, showcasing the versatility of anisotropic impedance surfaces. Although surface-to-leaky wave conversion has been well-established for isotropic metasurfaces, recent research has extended these principles to anisotropic structures, leading to improved performance. The study of holographic leaky-wave antennas with anisotropic impedance surfaces has gained significant attention due to their beam-steering capabilities and polarization diversity. Notable work in [4] demonstrated ultra-thin, high-efficiency holographic antennas capable of wide-angle beam steering, underscoring the potential of anisotropic modulation techniques. Contributions from Amini and Moeini in [5] advanced the field with wideband leaky-wave antennas using sinusoidally modulated anisotropic impedance surfaces, improving directivity and beam scanning performance. Additionally, the integration of wedge reflectors in two-dimensional holographic antennas [6] and the use of collimating cylindrical surface leaky waves [7] have enhanced radiation characteristics, bandwidth, and frequency-scanning capabilities. Recent research has explored temporally modulated anisotropic leaky-wave holograms. For instance, [8] proposed a theoretical framework to analyze the nonreciprocal behavior of temporally modulated anisotropic holograms, offering new beam-scanning mechanisms. The work in [9] extended these ideas by investigating spatio-temporally modulated impedance surfaces, offering fresh approaches to beam steering with anisotropic holographic designs. Despite these advancements, most designs focus on one-dimensional beam scanning in the elevation plane. Although wideband leaky-wave antennas with sinusoidally modulated impedance surfaces [5] and two-dimensional holographic antennas with wedge reflectors [6] achieve frequency-dependent beam steering, they remain largely constrained to unidirectional scanning. While cylindrical surface leaky waves enhance scanning in the elevation plane [7], they do not support azimuthal beam steering. Temporally modulated holograms [9] have primarily been developed for one-dimensional scanning, leaving a gap in the development of two-dimensional scanning solutions. As outlined in studies such as [10] and [11], antenna technologies are critical to meet the increasing bandwidth and capacity demands. Holographic antennas use holography principles to generate and direct beams toward specific targets. They achieve beam steering by manipulating surface impedance or meta-atomic states, allowing real-time beam steering without mechanical movement or complex electronics, making them ideal for dynamic beam control applications [12], [13]. Additionally, the integration of holographic MIMO surfaces in 6G wireless networks has been explored, offering both opportunities and challenges [14]. Recent works, such as [15] and [16], discuss the development of holographic antennas capable of one-dimensional beam steering, enhancing directional gain and reducing interference. The study in [17] highlights how varying surface impedance distributions manipulate wave polarization, improving performance across frequency bands. However, the need for two-dimensional beam steering is crucial for modern communication systems, as emphasized in [18] and [19]. These works underscore the need for antennas capable of dynamically adjusting their radiation patterns in both the azimuth and elevation planes. The design of antennas capable of circular polarization is particularly important for consistent performance across diverse environments. Research, including [20] and [21], has shown that antennas with axial ratios below 3 dB in the main lobe direction offer advantages in applications requiring robustness against polarization mismatch. Beam steering in the azimuth direction has been extensively studied, with phase-shifting techniques gaining prominence. The works of [22] and [23] demonstrate effective implementations of phased arrays to control beam directionality, achieving substantial signal quality improvements. In holographic antennas, beam steering is achieved through phase modulation and frequency variation. By applying phase shifts to the antenna illuminators, the azimuth direction can be precisely controlled. Additionally, frequency variation can steer the beam in the elevation direction, providing an efficient method for two-dimensional beam steering [13].

This paper is organized as follows. Section II presents the core design principles of a holographic antenna with anisotropic impedance surfaces, detailing its beam-steering capabilities in both azimuth and elevation directions, while maintaining circular polarization of leaky waves generated by the monopole array. Section III examines array design and surface implementation, deriving analytical expressions for the aperture field and far-field radiation pattern. These are influenced by three primary factors: the anisotropic impedance tensor, modulation coefficient, and complex propagation constant of the leaky wave. The section explains the beam-steering mechanism, covering azimuth steering through phase control and elevation steering via frequency adjustment, and includes a dispersion analysis to calculate the propagation constant based on impedance and modulation parameters. Section IV provides a detailed description of the holographic antenna's structure design, focusing on the modulation of surface impedance using a periodic array of subwavelength unit cells to enable controlled leaky wave propagation, and examines key design parameters including structure dimensions, modulation coefficient, pattern rotation, monopole spacing, and polarization. Section V presents a feeding network for the monopole array, along with its simulation results. Section VI details the practical implementation, including antenna fabrication using a Rogers 4003C substrate with modulated unit cells and a 1:16 SIW power divider for feeding the array. Finally, Section 0 discusses the design's advantages for 5G/6G applications, potential industrial applications, and directions for future research.

## II. DESIGN PROCEDURE OF A BEAM SCANNING HOLOGRAM

This section describes the methodology used to implement beam steering in the proposed antenna, which combines holographic theory with array theory, as shown in Fig. 1. This approach enables independent control of the radiation direction in both the azimuth ($\varphi$) and elevation ($\theta$) planes, while maintaining circular polarization to support consistent signal quality. The integration of holographic surface modulation (hologram) with array excitation (vertical monopoles) allows for practical beam steering and polarization control in a unified structure.

The antenna provides a peak gain of 24.5 dB and supports a fractional bandwidth of 22.22%, covering a wide frequency range suitable for high-data-rate applications. Circular polarization improves performance in environments with multipath propagation. The overall antenna size is 15 cm × 17 cm, offering a compact form factor relative to its performance. These specifications were selected based on system-level requirements and fabrication constraints.

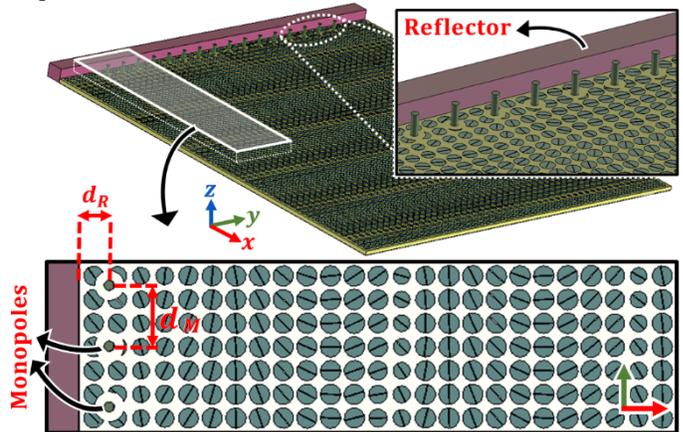

Fig. 1 presents the overall configuration of the proposed antenna, which synergistically combines holographic theory with array theory. The parameter $d_M = \lambda/2$ denotes the inter-monopole spacing, while $d_R = \lambda/4$ represents the separation between the monopole array and the reflector.

The design process began with analytical modeling to derive the surface impedance distribution required for beam steering and polarization control. MATLAB [24] was used to implement these models and to compute the initial excitation profiles and holographic patterns. The electromagnetic behavior of the antenna was then analyzed and optimized using full-wave simulations in COMSOL Multiphysics [25] and CST Microwave Studio [26]. These tools were employed to evaluate impedance matching, far-field characteristics, and polarization purity. Finally, a prototype was fabricated, and experimental measurements were conducted to validate the simulated results.

A. Beam Steering Mechanism

The antenna steers its beam in two dimensions using distinct techniques:

1) Elevation Steering (θ-direction):

Beam steering in the vertical plane is controlled by varying the operating frequency. As the frequency changes (within the 15–21 GHz band), the wavelength adjusts, altering the angle $\theta_0$ of the radiated beam. This frequency-based steering, rooted in holographic principles, is depicted in Fig. 2a, showing the top views of the holographic antenna varying the frequency induces rotation in the elevation plane ($\theta_0$). Fig. 2b presents a side view of the structure, highlighting the surface impedance ($Z_s$) and the directions of the tangential electric ($\vec{E_t}$) and magnetic ($\vec{H_t}$) fields, which together indicate the orientation of the surface current.

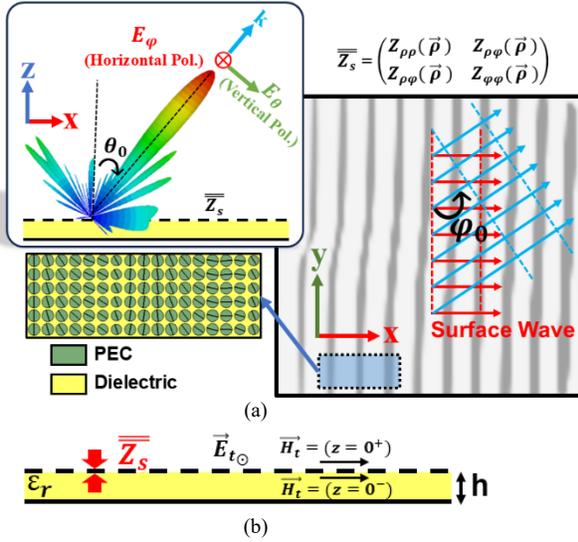

(a)

(b)

Fig. 2 (a) Top view illustrating azimuth steering by adjusting the progressive phase increment (Δβ) to control $\varphi_0$. (b) Side view showing elevation steering by varying frequency to adjust $\theta_0$.

2) Azimuth Steering (φ-direction):

Beam steering in the horizontal (azimuthal) plane is achieved based on array theory. The antenna utilizes an array of monopole feeds, each excited with a progressive phase increment denoted by $\Delta\beta$. By varying $\Delta\beta$, the main beam direction ($\varphi_0$) can be steered laterally, enabling dynamic azimuthal scanning.

This principle is illustrated in Fig. 2a, where adjusting Δβ leads to a corresponding shift in the azimuthal radiation angle.

This dual approach—phase control for azimuth and frequency variation for elevation—enables two-dimensional (2D) beam steering without mechanical movement, a key advantage for applications like autonomous vehicle tracking.

B. Circular polarization

The inset of Fig. 3 provides a detailed depiction of the holographic antenna, showcasing the local and overall electric field (E-field) distribution across a representative array of unit cells. the blue arrows represent the tangential E-field component, which we associate with $E_x = E_0 \cos(\omega t - kz)$, aligned along the x-direction in the antenna's plane. The yellow arrows represent the parallel E-field component, corresponding to $E_y = -E_0 \sin(\omega t - kz)$, along the y-direction, orthogonal to x. Both components are tangential to the antenna surface, as the plane lies in xy.

For right-hand circular polarization (RHCP), the far-field E-field is [27]:

$$\mathbf{E} = E_0(\hat{x}\cos(\omega t - kz) - \hat{y}\sin(\omega t - kz)) \quad (1)$$

This describes a wave propagating in +z, where:

- $E_x = E_0 \cos(\omega t - kz)$,
- $E_y = -E_0 \sin(\omega t - kz) = E_0 \cos(\omega t - kz + 90°)$,
- $E_y$ leads $E_x$ by 90°, and the vector rotates clockwise when viewed along +z, confirming RHCP.

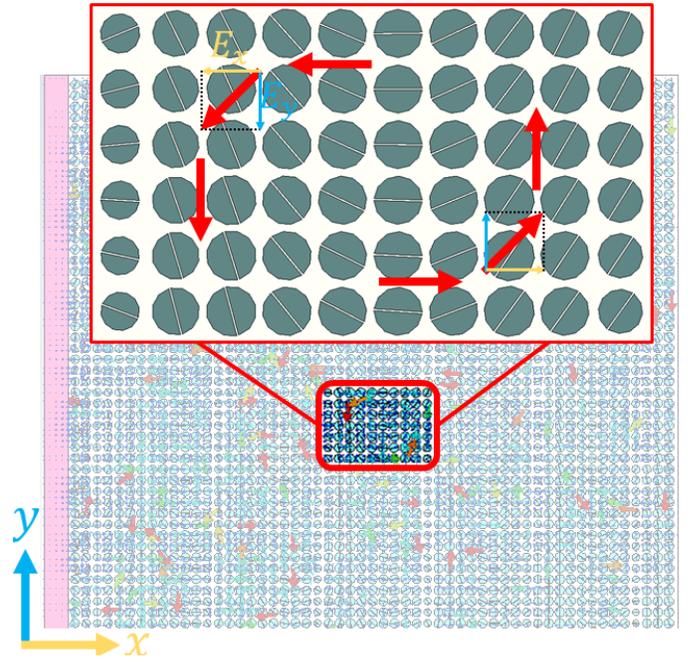

Fig. 3 Local and overall electric field (E-field) distribution across a representative array of unit cells within the holographic antenna structure, as obtained from CST simulations.

The red arrows likely illustrate the resultant E-field vector at different points on the antenna surface. In holographic antennas, the surface field distribution is engineered via impedance modulation to create a phase progression that, when radiated, produces this RHCP pattern. The spatial variation in the red arrows' directions reflects the phase differences between $E_x$ and $E_y$ across the structure, enabling the orthogonal components (blue and yellow arrows) to combine into the circularly polarized far-field wave.

III. ARRAY DESIGN AND SURFACE IMPLEMENTATION

The proposed antenna generates a steerable leakage wave to achieve precise beam steering in both elevation ($\theta_0$) and azimuth ($\varphi_0$) planes while maintaining circular polarization. The leakage wave results from the controlled interaction of a surface wave with an anisotropic impedance surface, modulated using holographic principles and array theory. This section outlines a systematic method to obtain the desired leakage wave, incorporating key relations to ensure accurate beam direction and polarization control.

A conceptual structure of anisotropic hologram is shown in Fig. 2a, comprising a modulated metasurface (Hologram) and an array of vertical monopoles acting as surface wave (SW) generators. These monopoles excite cylindrical magnetic surface waves that coherently superimpose to form a planar magnetic surface wave across the metasurface.

The tangential magnetic field at the surface, due to a single cylindrical source, can be approximated as [28]:

$$\vec{H}_t\big|_{z=0^+} \approx \vec{J}_{sw} H_1^{(2)}(k_{sw}\rho) \quad (2)$$

Where $\vec{H}_t$ is the tangential magnetic field vector at $z = 0^+$ represents the surface current, $H_1^{(2)}$ denotes the Hankel function of the second kind (first order), $k_{sw} \triangleq k^{(0)} = \beta - j\alpha$ is the complex surface wave number, and $\rho$ is the radial distance in the surface plane.

For an array of N=16 monopole uniformly spaced at $d_M = \lambda/2$, the total surface magnetic field at an observation point $\vec{r} = (x, y, z = 0^+)$ is given by:

$$\vec{H}_t\big|_{z=0^+} \approx \vec{J}_{sw} \sum_{n=1}^{N} H_1^{(2)}(k_{sw}\rho_n)e^{j\phi_n} \quad (3)$$

Where:
- $\rho_n = \sqrt{(x-x_n)^2 + y^2}$ is the distance between the observation point and the n-th monopole,
- $x_n = \left(n - \frac{N+1}{2}\right), d_M = (n - 7.5)\lambda/2$ is the x-position of the n-th monopole (all located along y=0, z=0).

To steer the resulting surface wave in the desired azimuthal direction $\varphi_0$, each monopole is fed with a phase shift given by:

$$\phi_n = -kd_M n\sin(\varphi_0) \quad (4)$$

Substituting $k = \frac{2\pi}{\lambda}$ and $d_M = \lambda/2$, this simplifies to:

$$\phi_n = -\pi n \sin(\varphi_0) \quad (5)$$

Defining $\Delta\beta \triangleq -\pi\sin(\varphi_0)$, the phase shift becomes:

$$\phi_n = n\Delta\beta \quad (6)$$

$\vec{J}_{sw}$: constant surface current vector (same for all sources).
$\vec{H}_t$: tangential vector magnetic field at z=0$^+$.
$H_1^{(2)}$: Hankel function of the second kind, first order.
$k_{sw} \triangleq k^{(0)} = \beta - j\alpha$: complex surface wave number.
n = 0, 1, ... ,N−1: Monopole element index.
$d_M = \lambda/2$: Inter-element spacing.
$\rho_m$: Distance from observation point to the n-th monopole.
$\varphi_0$: desired Azimuth beam steering angle.
$\phi_n$: phase shift applied to the n-th monopole.
$\Delta\beta$: progressive phase increment.

In the azimuth plane, the angle $\phi_0$ is controlled by applying a progressive phase increment $\Delta\beta$ to the monopole feed array, as illustrated in Fig. 2a. Dynamic beam steering is achieved by exciting each feed element with a complex phase excitation of exponential form ($e^{jn\Delta\beta}$), where $n$ is the element index and $\Delta\beta$ is the progressive phase increment. Discrete values of $\Delta\beta = \{-100°, -50°, 0°, 50°, 100°\}$ are applied to the monopole array in the x-y plane using COMSOL simulations to demonstrate the azimuth steering capability, as illustrated in Fig. 4. COMSOL significantly reduces simulation run time.

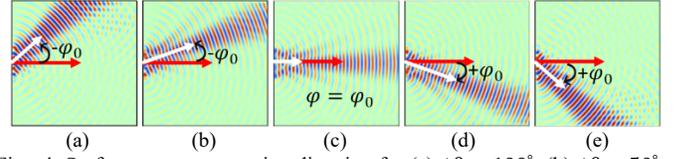

Fig. 4. Surface wave propagation direction for (a) Δβ=−100°, (b) Δβ=−50°, (c) Δβ=0°, (d) Δβ=50°, (e) Δβ=100°. Red and white arrows indicate the x-direction and wave vector direction, respectively.

In the elevation plane, the angle $\theta_0$ is adjusted by varying the operating frequency within the 16–20 GHz band, which alters the wavelength and holographic pattern phase, as depicted in Fig. 2a.

The metasurface consists of semi-regularly arranged patches printed on a dielectric substrate backed by a ground plane (as shown in Fig. 2b. When the dimensions of these patches are much smaller than the wavelength (approximately λ/6), the structure behaves as an effective impedance surface. By precisely engineering the impedance distribution, it is possible to convert a surface wave into a leaky wave with the desired propagation direction, polarization state, and topological charge.

### A. Surface impedance distribution

In the analysis of leaky-wave holograms, the characteristics of metasurface can be described by the impedance boundary condition. For the metasurface placed at z = 0 (see Fig. 2b), the transparent impedance boundary condition can be expressed as [29]:

$$\vec{E}_t = j\bar{\bar{X}}(\vec{\rho}).\hat{z}\left(\vec{H}_t\big|_{z=0^+} - \vec{H}_t\big|_{z=0^-}\right) = j\bar{\bar{X}}(\vec{\rho}).\vec{J} \quad (7)$$

Where,

$$\bar{\bar{X}}(\vec{\rho}) = \begin{pmatrix} X_{\rho\rho}(\vec{\rho}) & X_{\rho\varphi}(\vec{\rho}) \\ X_{\rho\varphi}(\vec{\rho}) & X_{\varphi\varphi}(\vec{\rho}) \end{pmatrix} \quad (8)$$

The tensorial reactance, denoted as $\bar{\bar{X}}(\vec{\rho})$, varies with the observation vector $\vec{\rho}$ in cylindrical coordinates. The tangential electric and magnetic fields are represented by $\vec{E}_t$ and $\vec{H}_t$, respectively. Typically, the surface reactance distribution can be described as derived from the generalized holographic theory [30]:

$$X_{\rho\rho}(\vec{\rho}) = X_\rho\left[1 + m_\rho(\vec{\rho})\cos\left(Ks(\vec{\rho}) + \Phi_\rho(\vec{\rho})\right)\right] \quad (9)$$

$$X_{\rho\varphi}(\vec{\rho}) = X_\rho m_\varphi(\vec{\rho})\cos\left(Ks(\vec{\rho}) + \Phi_\rho(\vec{\rho})\right) \quad (10)$$

$$X_{\varphi\varphi}(\vec{\rho}) = X_\varphi\left[1 - m_\rho(\vec{\rho})\cos\left(Ks(\vec{\rho}) + \Phi_\rho(\vec{\rho})\right)\right] \quad (11)$$

In equations (9)-(11), the coefficients $X_\rho$ and $X_\varphi$ represent the average surface reactances, which do not depend on the position vector $\vec{\rho}$. Additionally, $m_\rho$ and $m_\varphi$ are the modulation indices that govern the leakage constant distribution across the radiation aperture. It should be noted that $Ks(\vec{\rho})$ is the fast-varying component, while $\Phi_{\rho,\varphi}(\vec{\rho})$ represents the slow-varying components of the modulation phase. Equation (9) through (11) are computed using MATLAB.

## B. Calculation of far-zone field

A widely used approach for designing leaky-wave holograms involves the aperture field estimation technique [31, 32]. This method describes the relationship between the surface reactance tensor and the aperture field vector ($\vec{E}_{ap}$) as follows [32]:

$$\bar{\bar{X}}(\vec{\rho}).\hat{\rho} = X_0[\hat{\rho} + 2Im\{\frac{\vec{E}_{ap}}{-J_{sw}H_1^{(2)}(k^{(0)}\rho)}\}] \quad (12)$$

Here, $H_1^{(2)}$ represents the surface wave function excited by the monopole array launcher. The aperture field vector can be expressed as a combination of its x and y components.

$$\vec{E}_{ap}(\vec{\rho}) = E_{ax}(\vec{\rho})\hat{x} + E_{ay}(\vec{\rho})\hat{y} \quad (13)$$

To evaluate the aperture field vector, three key factors need to be addressed: 1) The field magnitude distribution, which shapes the radiated beam; 2) The field phase, which governs the beam's direction and vorticity characteristics; and 3) The interplay between the x and y components of the aperture field, which defines the polarization of the emitted wave. The aperture field vector components, as outlined in [33], are defined as follows:

$$E_{ax,ay}(\vec{\rho}) \quad (14)$$
$$= M_{x,y}(\vec{\rho}) \frac{J_{SW}}{\sqrt{2\pi\rho\sqrt{[\beta_{UM}^{(0)} + \delta\beta(\vec{\rho})]^2 + \delta\alpha^2(\vec{\rho})}}}$$
$$\times e^{-\delta\alpha(\vec{\rho})\rho}e^{-j[k\rho sin\theta_0 cos(\phi-\phi_0)+l\phi]}$$

Here, $M_x$ and $M_y$ represent modulation indices, defined as M cos φ and M sin φ, respectively. $\beta_{UM}^{(0)}$ denotes the unmodulated propagation constant, representing the baseline phase propagation constant of a surface wave in an isotropic medium. This serves as a reference for understanding wave propagation characteristics before considering surface-induced perturbations. The term $\delta\beta(\vec{\rho})$ accounts for spatially varying perturbations in the propagation constant due to the anisotropic nature of the surface. These perturbations arise from factors such as surface roughness, material heterogeneity, and directional dependence of wave speed, leading to modifications in the wave's phase velocity as it propagates over the surface. Similarly, $\delta\alpha(\vec{\rho})$ represents spatially dependent variations in the attenuation constant caused by the anisotropic surface. These variations influence the wave's energy dissipation as it propagates, incorporating effects like surface roughness and material heterogeneity into the model. Together, these terms are incorporated into the aperture field equations to accurately model the behavior of surface waves on anisotropic surfaces, ensuring both phase and attenuation characteristics are appropriately accounted for.

In examining radiative apertures, Fourier transformation and stationary phase theory are utilized to calculate the far-zone fields. Consequently, the θ and φ components of the far-zone field are obtained as follows [27]:

$$\tilde{F}_\theta(\theta, \varphi) = \tilde{E}_{ax}\cos\phi + \tilde{E}_{ay}\sin\phi \quad (15)$$
$$\tilde{F}_\phi = \cos\theta\,(-\tilde{E}_{ax}\sin\phi + \tilde{E}_{ay}\cos\phi) \quad (16)$$

where $\tilde{E}_{ax}$ and $\tilde{E}_{ay}$ represent the Fourier transforms of $E_{ax}$ and $E_{ay}$, respectively. Thus, in the cylindrical coordinate system, we can express:

$$\tilde{E}_{ax} = \iint_{ap} E_{ax} e^{jk\rho sin\theta cos(\phi-\acute{\phi})}\rho d\rho\,d\acute{\phi} \quad (17)$$
$$\tilde{E}_{ay} = \iint_{ap} E_{ay} e^{jk\rho sin\theta cos(\phi-\acute{\phi})}\rho d\rho\,d\acute{\phi} \quad (18)$$

## C. Design and analysis of hologram

To verify the analytical method presented in the previous section, an anisotropic hologram with circular polarization was designed. The antenna operating frequency was set at 18 GHz for general applicability. The substrate utilized is Rogers RO4003, characterized by a dielectric constant of 3.55, a loss tangent (tan δ) of 0.0027, and a thickness of 1.524 mm. An anisotropic impedance was realized using a circular patch featuring a central slot, printed on the grounded dielectric substrate. The unit cell period was selected as 2.9 mm, corresponding to one-sixth of the wavelength (λ/6) at the operating frequency. Fig. 5a shows the proposed anisotropic unit cell. the diameter of the circle parameter (denoted as d) and its orientation angle (namely ψ) are considered for changing the impedance, obtained through eigenmode analysis using CST Microwave Studio. To extract the impedance tensor, first the reactance's $X_1$ and $X_2$ are calculated for the surface wave propagating along the x and y directions, respectively.

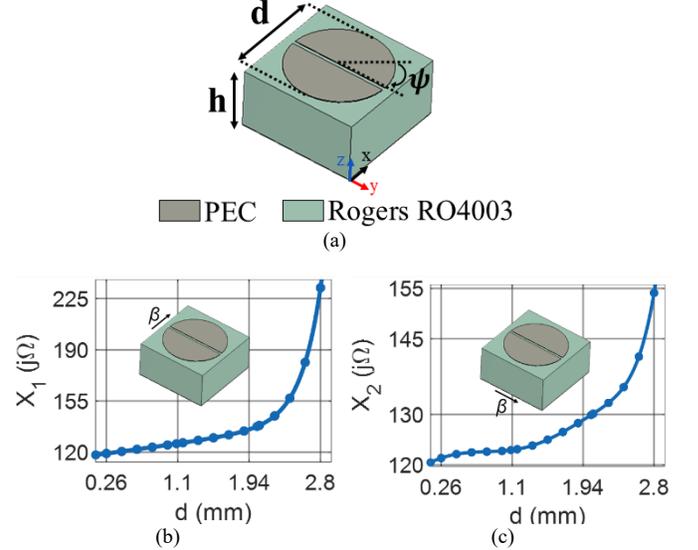

Fig. 5. (a) Asymmetric circular patch for realization of tensorial impedance. (b) Impedance curve for the wave propagating along the x axis ($X_1$). (c) Impedance curve for the wave propagating along the y axis ($X_2$).

In both directions, the parameter ψ is kept equal to zero and the parameter d is varied from 0.26 to 2.8 mm. The reactance curves for $X_1$ and $X_2$ are plotted in Fig. 5b and Fig. 5c, respectively. For the retrieval of the impedance tensor the method proposed in [34] is used, which can be expressed as:

$$\bar{\bar{Z}}_s = j\bar{\bar{X}}_s = jR^T(\psi)\bar{\bar{X}}(d)R(\psi) \quad (19)$$

Where,

$$\bar{\bar{X}}(d) = \begin{pmatrix} X_1(d) & 0 \\ 0 & X_2(d) \end{pmatrix} \quad (20)$$

And R is the rotation matrix:

$$R(\psi) = \begin{pmatrix} \cos(\psi) & -\sin(\psi) \\ \sin(\psi) & \cos(\psi) \end{pmatrix} \quad (21)$$

Note that T indicates the transpose operator. Fig. 6 shows the impedance maps of proposed anisotropic unit cell versus the diameter and orientation angle of patch.

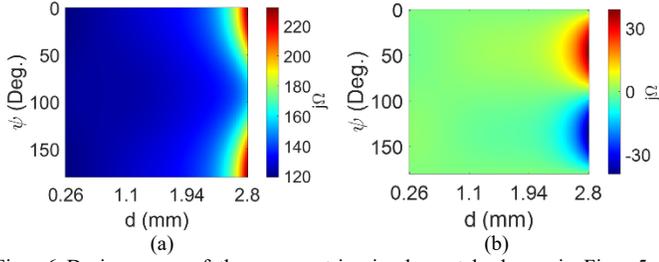

Fig. 6 Design maps of the asymmetric circular patch shown in Fig. 5a, illustrating the variation with diameter and orientation angle: (a) $X_{\rho\rho}$. (b) $X_{\rho\varphi}$.

### D. Aperture fields for anisotropic holograms

in anisotropic holograms the components of field vector can be defined independently. This enables us to control the polarization of radiation field. To achieve circular polarization at $(\theta_0, \varphi_0)$ the horizontal and vertical components of far-zone field must satisfy the following condition [33]:

$$\tilde{F}_\phi(\theta_0, \varphi_0) = \pm j\tilde{F}_\theta(\theta_0, \varphi_0) \quad (22)$$

where signs + and - represent the left-hand and righthand polarizations, respectively. Equation (22) imposes the following condition on aperture field components:

$$E_{ay}(\vec{\rho}) = E_{ax}(\vec{\rho}) \frac{cos\theta_0\, sin\phi_0 + e^{\pm j\pi/2}\, cos\phi_0}{cos\theta_0\, cos\phi_0 - e^{\pm j\pi/2}\, sin\phi_0} \quad (23)$$

If $E_{ax}$ is defined according to Equation (14), the modulation indices must have the following relationship:

$$M_y(\vec{\rho}) = M_x(\vec{\rho}) \frac{cos\theta_0\, sin\phi_0 + e^{\pm j\pi/2}\, cos\phi_0}{cos\theta_0\, cos\phi_0 - e^{\pm j\pi/2}\, sin\phi_0} \quad (24)$$

To have radiation with right circular polarization (RHCP) in the broadside $(\theta_0 = 0°$ and $\varphi_0 = 0°)$, the following condition must be met:

$$M_y(\vec{\rho}) = -jM_x(\vec{\rho}) = -jM \quad (25)$$

where M is assumed to be constant. Using (15) and (16) we can estimate the far-zone components as:

$$\tilde{F}_\theta(\theta, \phi) \quad (26)$$
$$= \iint_{ap} \frac{MJ_{SW} e^{-j\phi}}{\sqrt{2\pi\rho \sqrt{\left[\beta_{UM}^{(0)} + \delta\beta(\vec{\rho})\right]^2 + \delta\alpha^2(\vec{\rho})}}}$$
$$\times e^{-\delta\alpha(\vec{\rho})\rho} e^{jk\acute{\rho}sin\theta\, cos(\phi-\acute{\phi})} e^{-jl\acute{\phi}} \acute{\rho} d\acute{\rho} d\acute{\phi}$$

$$\tilde{F}_\phi(\theta, \phi) \quad (27)$$
$$= \iint_{ap} \frac{-jMJ_{SW} e^{-j\phi} cos\theta}{\sqrt{2\pi\rho \sqrt{\left[\beta_{UM}^{(0)} + \delta\beta(\vec{\rho})\right]^2 + \delta\alpha^2(\vec{\rho})}}}$$
$$\times e^{-\delta\alpha(\vec{\rho})\rho} e^{jk\acute{\rho}sin\theta\, cos(\phi-\acute{\phi})} e^{-jl\acute{\phi}} \acute{\rho} d\acute{\rho} d\acute{\phi}$$

Fig. 7 shows the synthesized surface impedance distribution to generate right-hand polarization.

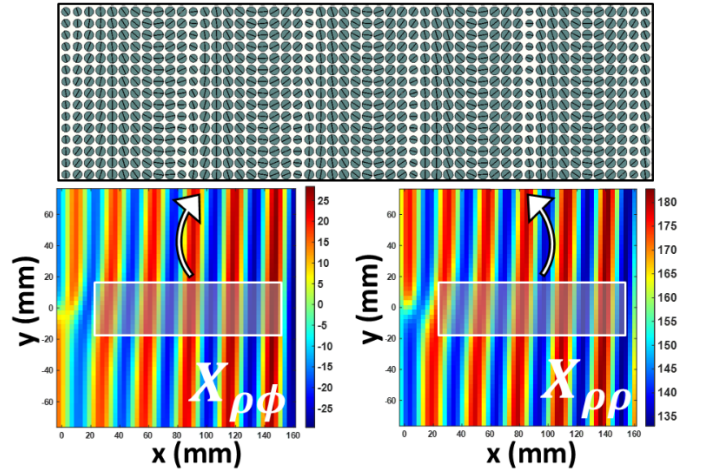

Fig. 7 Anisotropic surface impedance distribution and realized model.

The axial ratio is calculated to confirm polarization quality [27]:

$$AR = 20\log_{10}(\frac{|\widetilde{F_\theta}| + |\widetilde{F_\phi}|}{|\widetilde{F_\theta}| - |\widetilde{F_\phi}|}) \quad (28)$$

An axial ratio ≤ 4 dB ensures effective circular polarization for applications such as 5G/6G networks and autonomous vehicle tracking.

This method provides a robust framework for generating a leakage wave with precise beam steering and circular polarization, leveraging the antenna's holographic surface and array excitation. The incorporated equations ensure accurate synthesis of the leakage wave, validated through simulations and measurements, making the antenna suitable for high-data-rate applications.

The design is implemented by creating an impedance profile with patterned unit cell patches, optimized to align with the synthesized impedance tensor. Full-wave simulations in COMSOL Multiphysics and CST Microwave Studio optimize the leakage wave's characteristics, confirming a peak gain of 24.5 dB and a half-power beamwidth of 8.7° at 18 GHz. A prototype is fabricated, and experimental measurements validate the leakage wave's beam direction, gain, and polarization, ensuring alignment with design specifications.

As depicted in Fig. 8, The antenna design process integrates COMSOL Multiphysics, CST Studio Suite, and MATLAB to create an efficient, high-performance antenna through iterative simulation and optimization. A 2D surface is first defined in COMSOL, while progressive phase increment on 16 monopoles evaluate pattern steerability. Unit cell parameters are chosen in CST to calculate the dispersion curve, and MATLAB computes surface impedance to arrange unit cells into an antenna, with its size determined. CST simulations verify gain and side lobe levels; if unmet, the antenna size is adjusted in MATLAB and re-simulated. Upon success, a power divider is designed in CST, and coupling methods (coaxial cable, SMA connector, slot) are simulated to select the most cost-effective option. The process culminates in fabrication and measurement, ensuring a robust antenna design validated through simulation and testing.

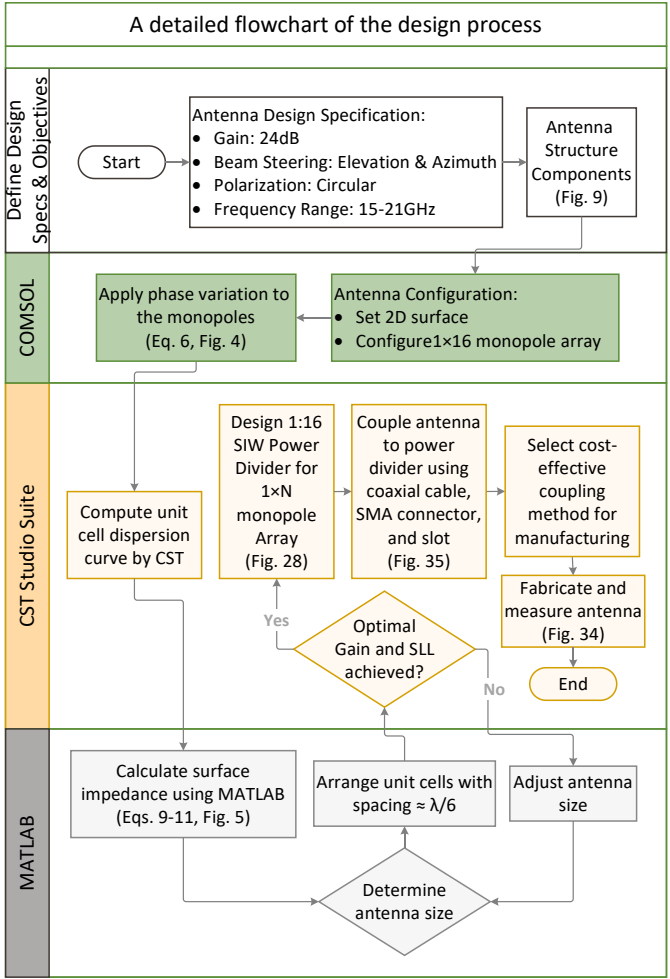

Fig. 8. Flowchart of the Design Process.

## IV. STRUCTURE DESIGN

A holographic antenna employs a periodic array of subwavelength unit cells to modulate the surface impedance, thereby enabling controlled excitation and propagation of leaky waves across the structure. In the proposed design, each unit cell consists of a circular metallic patch with a central slot, patterned on a Rogers 4003C substrate, which has a relative permittivity of 3.55 and a thickness of 1.524 mm (60 mil). This configuration facilitates low-loss wave propagation and high-efficiency radiation performance. The complete antenna structure, depicted in Fig. 9, comprises a ground-backed reflector, an array of monopole feed, patterned patches, the Rogers 4003C dielectric substrate, a continuous ground plane, and a coaxial feeding network.

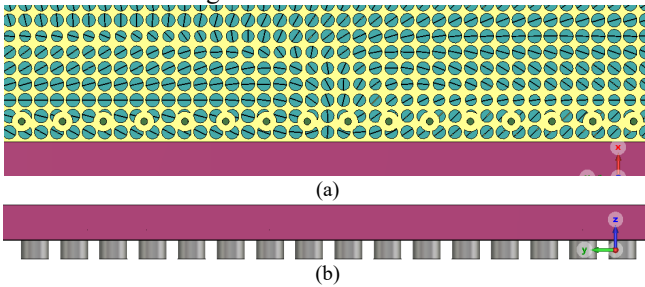

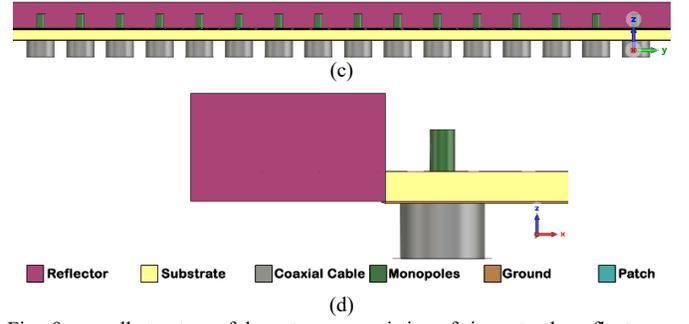

Fig. 9. overall structure of the antenna, consisting of six parts: the reflector, substrate, coaxial cable, monopoles, ground and patch (holographic surface) views (a) top, (b) back, (c) front, (d) side.

In the following, we will explore the details of the design, including the structure dimensions, modulation coefficient, number of monopoles, pattern rotation, monopole spacing, polarization, and the behavior of the anisotropic structure.

### A. Structure Dimensions

As seen in Fig. 1, the individual cells are modulated on the surface and appear to be repeating in the x-direction. As the wave propagates along the structure, a portion of the electromagnetic radiation gradually leaks out into space. Eventually the wave will completely leak out of the structure and will die out after travelling some distance. If the structure extended in the x-direction to infinity, its radiation into space cannot be sustained indefinitely. Clearly, increasing the linear length of structure in the x-direction will not necessarily increase the antenna gain. Consequently, we need to find the optimal dimensions of the antenna structure. In the following, the first estimates of structure dimensions will be examined.

The optimal antenna dimensions were determined through analytical methods and validated using computational simulations for comparison. The results are also verified by observing the surface currents. The directivity is given by [27]:

$$D = 10\log\left(\frac{4\pi A}{\lambda^2}\right) \quad (29)$$

where A is the physical aperture area of the antenna and λ is the wavelength. At the center frequency of 18 GHz (λ=16.67mm), three different hologram sizes were evaluated to assess their impact on gain. Table. I summarizes the calculated directivities for these configurations. This analysis guided the selection of the 15×17cm design (Case 2), which offers a realized gain of 23.1 dB and demonstrates stable performance under azimuthal rotation, as illustrated in Fig. 10b.

Table. I Comparison of directivity and other antenna characteristics for different hologram dimensions.

| | Dimension $(cm)^2$ | Number of Monopoles | Theoretical Gain (dB) | Realized Gain (dB) | Rad. Efficiency |
|---|---|---|---|---|---|
| 1 | 7×17 | 8 | 27.3 | 20 | %85 |
| 2 | 15×17 | 16 | 30 | 23.1 | %86 |
| 3 | 30×30 | 24 | 36 | 27 | %85 |

In Table. I the length, width and area A of antenna are given. The 3D radiation pattern of three cases is illustrated in Fig. 10.

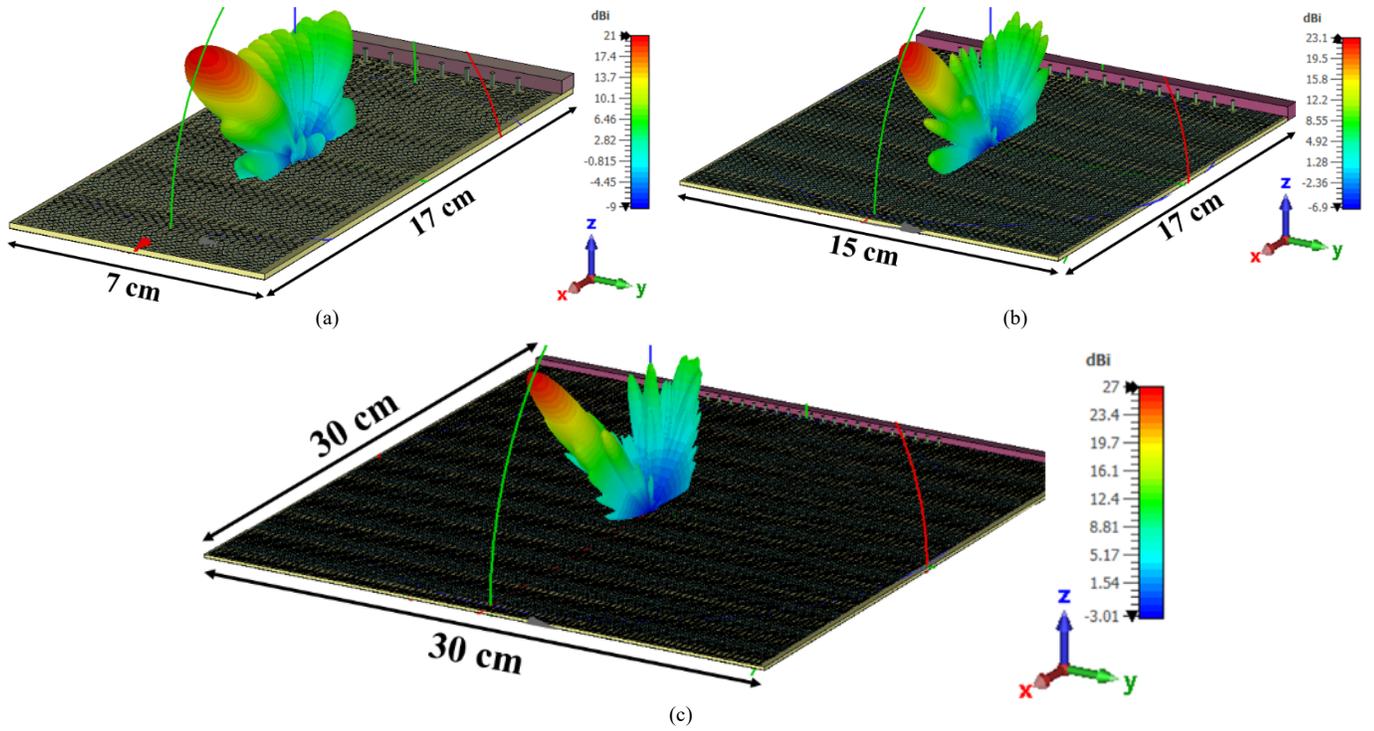

Fig. 10. 3D pattern and gain for dimensions listed in Table. I.

Fig. 10a illustrates that Case 1's 3D radiation pattern achieves a gain of about 21 dB, but it is highly sensitive to slight rotations in the φ-direction, leading to significant pattern distortions. To address this, Case 2 increases the structure's y-direction dimension, stabilizing the gain against φ-directional rotations and making it a robust design option. Fig. 10c presents the results for Case 3, while Fig. 11 shows its surface current distribution. The region within the black line indicates where surface currents propagate up to a limit, beyond which current density drops sharply. This suggests that the antenna's radiation mechanism extends roughly halfway to the specified dimensions before radiating. Limiting the antenna length to the black line (x ≈ 15 cm) should yield acceptable efficiency.

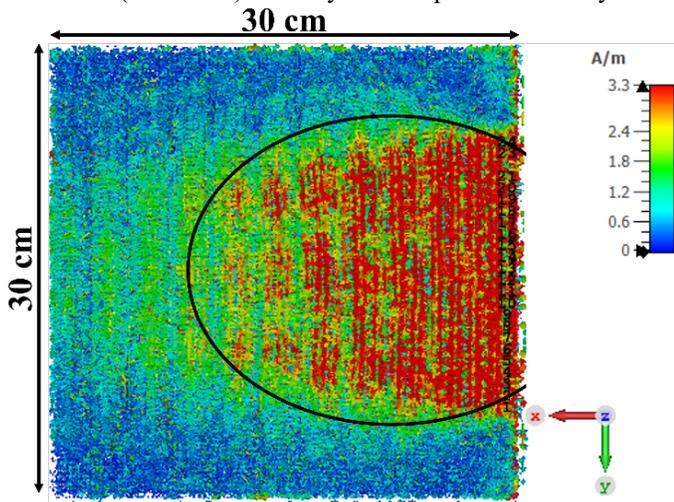

Fig. 11. Surface current distributions for case 3.

Choosing between Cases 2 and 3 involves balancing higher gain against lower cost, requiring a careful design approach. A detailed analysis of pattern rotation is provided in Section C.

### B. Modulation Coefficient

As illustrated in Fig. 1, the unit cells of the structure are periodically arranged along the x-axis with a periodicity determined by the modulation coefficient M, as defined in equation (14). This coefficient directly influences the amplitude of the electric field and sets the repetition period of the unit cells in the holographic structure. By varying M, its impact on the antenna's gain and side lobe level is analyzed, as shown in Fig. 12. The amplitude of the aperture can be controlled to some extent by manipulating the modulation depth, while polarization may be controlled by the use of a tensor formulation [35].

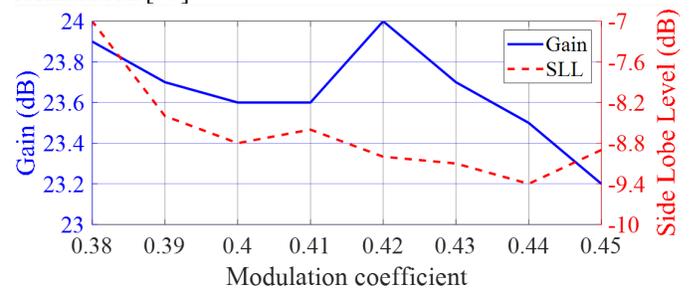

Fig. 12. Modulation coefficient of the structure relative to gain and side lobe level for frequency 18 GHz.

### C. Pattern Rotation

Fig. 13 shows the antenna monopole array labels from right to left, and application of applying progressive phase increment ($\Delta\beta$) to each one.

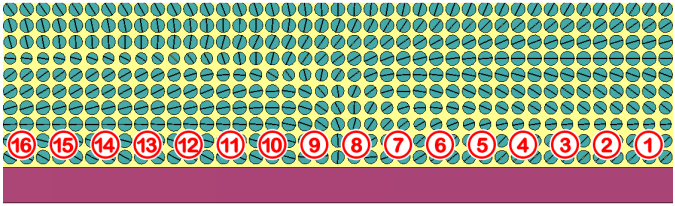

Fig. 13. Configuration of monopole elements indexed from $n = 1\ to\ 16$, showing the corresponding applied phase distribution.

In fact, two general cases can occur: $\Delta\beta = 0$ or $\Delta\beta \neq 0$. We will examine both cases.

1) *Case 1: $\Delta\beta = 0$ (Elevation beam steering)*

When $\Delta\beta = 0$, all monopoles are excited in phase, resulting in a radiation pattern directed along $\varphi = 0°$. Fig. 14 displays the antenna's elevation beam steering, showing two-dimensional patterns in the $\varphi = 0°$ plane for frequencies from 15 to 21 GHz. The objective is to evaluate how antenna gain and radiation direction vary in the $\theta$-direction when the frequency shifts by ±1, ±2, or ±3 GHz from the center frequency. To achieve this, the antenna patterns are analyzed in Cartesian coordinates across the 15 to 21 GHz range, as shown in Fig. 14.

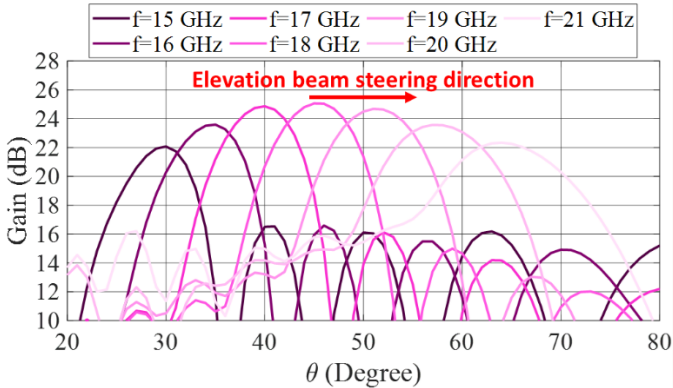

Fig. 14. Two-dimensional pattern in the cartesian coordinates for frequencies from 15 to 21 GHz in the $\varphi = 0°$ plane.

Table. II lists the antenna gain and radiation direction in the $\varphi = 0°$ plane, half-power beamwidth, and side lobe levels for frequencies from 15 to 21 GHz.

Table. II comparison of the antenna characteristics for different frequencies.

| Freq. (GHz) | Gain (dB) | SLL in $\varphi = 0°$ plane (dB) | HPBW in $\varphi = 0°$ plane (deg.) | $\theta_0$ | SLL in $\theta = \theta_0$ plane (dB) | HPBW in $\theta = \theta_0$ plane (deg.) |
|---|---|---|---|---|---|---|
| 15 | 21.7 | -5.5 | 7.4 | 30 | -9.8 | 16.2 |
| 16 | 23.3 | -7 | 8 | 35 | -11.6 | 13.3 |
| 17 | 24.5 | -8.7 | 8.1 | 40 | -13 | 11.4 |
| 18 | 24.5 | -10.1 | 8.7 | 45 | -12.9 | 9.6 |
| 19 | 23.9 | -9.6 | 9.7 | 51 | -10.6 | 8.2 |
| 20 | 22.6 | -8.3 | 11.7 | 57 | -11.4 | 7.2 |
| 21 | 21.2 | -6.1 | 13.5 | 64 | -14.7 | 6.7 |

Fig. 15 illustrates the three-dimensional radiation patterns of the antenna from a side view at frequencies of 17, 18, and 19 GHz, highlighting pattern rotation in the elevation plane.

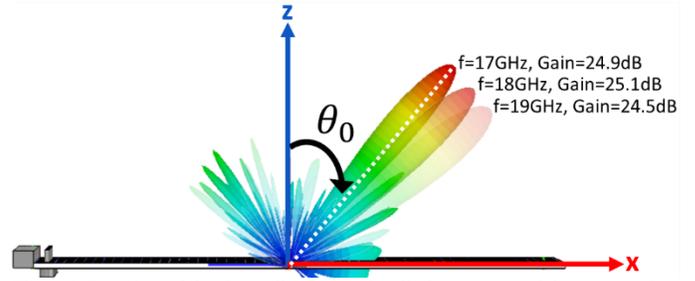

Fig. 15. Rotation of the three-dimensional radiation pattern of the antenna in the elevation plane for frequencies 17 GHz, 18 GHz and 19 GHz.

The beam rotation angle in the $\theta$-direction is analyzed to observe its variation with frequency, as shown in Fig. 16 for 17, 18, and 19 GHz. The beam angle is affected by the phase difference applied to the monopoles, with the ideal zero-slope condition indicated by the dotted line.

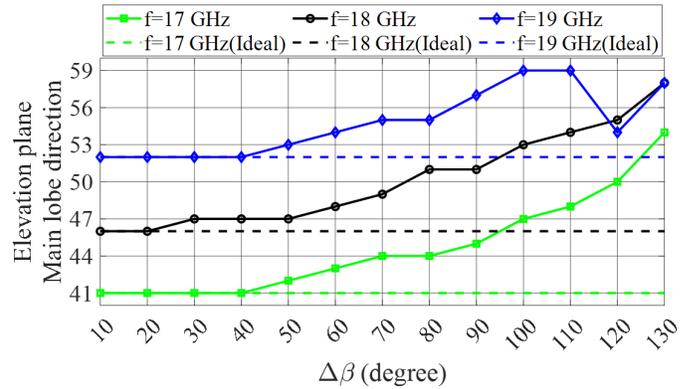

Fig. 16. Error due to applying phases to monopoles causing beam elevation drop for frequencies 17, 18, and 19 GHz.

Changes in the direction of radiation beam in the elevation plane as a function of frequency change and phase shift ($\Delta\beta$) applied to the monopoles.
Fig. 16 illustrates that the beam rotation angle in the Elevation plane is influenced by phase variations applied to the monopoles. The maximum effect observed is 13°, 12°, and 7° at frequencies of 17 GHz, 18 GHz, and 19 GHz, respectively. Additionally, Fig. 16 shows the impact of phase shifts on the antenna beam angle within the Elevation plane half-power beamwidth (HPBW). The HPBW in the Elevation plane widens by approximately 6° for every 1 GHz increase in frequency, indicating that larger phase shifts applied to the monopoles result in a broader Elevation plane HPBW.

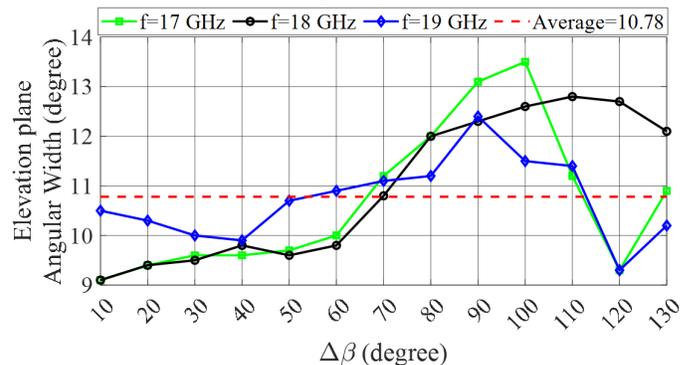

Fig. 17. The sensitivity of the beamwidth in the Elevation plane direction to the change in phase applied to the monopoles for frequencies 17, 18, and 19 GHz.

Now we examine the results for the beamwidth in the θ-direction and find the related suitable beamwidth in the φ-direction. The reason for selecting the beamwidth in the elevation direction is that, as shown in Fig. 16, the beamwidth in the θ-direction is insensitive to the variations of applied phases to monopoles, and also frequency. It is almost always constant.

To determine the optimal number of monopoles, two key factors must be considered: First, the structure should be excited by a plane wavefront, requiring a sufficient number of monopoles to effectively form such a front. Second, as phase shifters typically provide $2^n$ bit resolution, the number of monopoles should correspond to powers of two (e.g., N = 2, 4, 6, 8, 16, 32). By averaging the beamwidths in the Elevation plane, as shown in Fig. 17, a value of 10.78 degrees is obtained. Optimizing the half-power beamwidth (HPBW) in both the Azimuth and Elevation directions is crucial for achieving the best resolution in a pencil-beam pattern.

As shown in Fig. 18, the average HPBW in the Azimuth plane across the frequencies of 17, 18, and 19 GHz is approximately 11.09 degrees, closely matching the average HPBW in the Elevation plane. This balance is achieved by selecting N=16, where N denotes the number of monopoles.

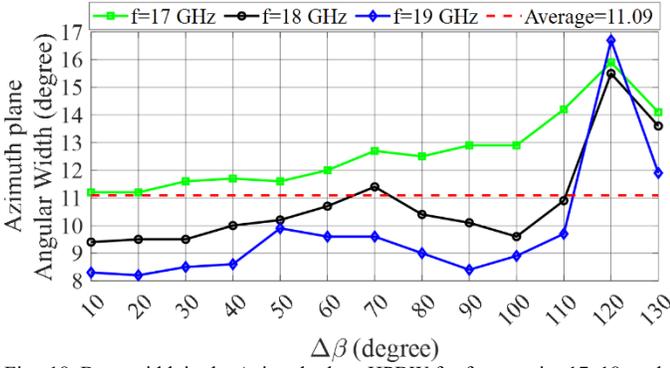
Fig. 18. Beamwidth in the Azimuth plane HPBW for frequencies 17, 18, and 19 GHz with N=16.

The circular and elliptical beam shapes in Fig. 19 further demonstrate that choosing N=16 results in nearly equal HPBWs in both the Azimuth and Elevation planes.

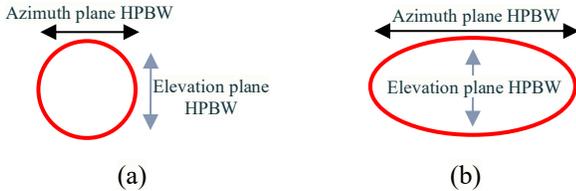
Fig. 19. HPBW for (a) N = 16 and (b) N < 16, where N is the number of monopoles.

The pattern rotation in the θ-direction has been analyzed. Subsequently, the pattern rotation in the φ-direction will be investigated.

*2) Case 2: $\Delta\beta \neq 0$ (Azimuth beam steering)*

In this case, each monopole is excited with a distinct phase, directing the radiation beam toward $\varphi = \varphi_0$, as determined by Equation (6). For $\Delta\beta \neq 0$, the pattern in the φ-direction will rotate. The following phase differences ($\Delta\beta$) are applied: Monopole 1 receives a phase of 0 degrees, Monopole 2 receives $\Delta\beta$ degrees, Monopole 3 receives $2\Delta\beta$ degrees, Monopole 4 receives $3\Delta\beta$ degrees, up to Monopole 16, which receives a phase of $15\Delta\beta$ degrees. The antenna simulation has been performed for $\Delta\beta = 108$, and the three-dimensional pattern is shown in Fig. 20.

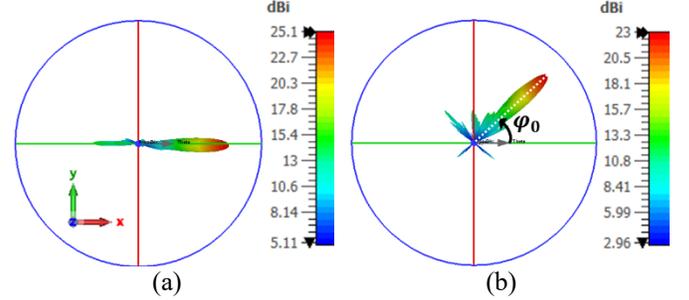
Fig. 20. Three-dimensional pattern rotation of the far-field of the antenna in the Azimuth plane due to the phase difference application. (a) $\Delta\beta = 0$ and (b) $\Delta\beta = 108$.

Table. III Comparison of antenna performance for various cases at the center frequency of 18 GHz.

| $\Delta\beta$ (deg.) | Gain (dB) | $\varphi_0$ | SLL in $\varphi = \varphi_0$ plane (dB) | HPBW in $\varphi = \varphi_0$ plane (deg.) | $\theta_0$ | SLL in $\theta = \theta_0$ plane (dB) | HPBW in $\theta = \theta_0$ plane (deg.) |
|---|---|---|---|---|---|---|---|
| 0 | 24.5 | 0 | -10 | 8.6 | 46 | -10 | 9.5 |
| 10 | 24.5 | 4 | -10.2 | 8.7 | 46 | -10.3 | 9.4 |
| 20 | 24.4 | 8 | -10.4 | 8.7 | 46 | -11.4 | 9.5 |
| 30 | 24.2 | 12 | -10.7 | 8.8 | 46 | -12.2 | 9.6 |
| 40 | 24.1 | 17 | -11.7 | 8.9 | 47 | -11.3 | 10 |
| 50 | 23.9 | 21 | -12 | 9 | 47 | -11.1 | 10.3 |
| 60 | 23.6 | 27 | -12 | 9.3 | 49 | -12.2 | 10.7 |
| 70 | 23.3 | 31 | -13 | 9.5 | 49 | -12.3 | 10.5 |
| 80 | 23.2 | 36 | -15.9 | 10.3 | 51 | -11.8 | 9.7 |
| 90 | 22.8 | 40 | -15.3 | 10.6 | 52 | -10.8 | 9.3 |
| 100 | 22.3 | 44 | -16.4 | 11.1 | 53 | -12 | 8.8 |
| 110 | 21.2 | 48 | -13 | 11.7 | 55 | -12.9 | 9.6 |
| 120 | 19.8 | 52 | -13 | 12.4 | 56 | -10.5 | 10.2 |

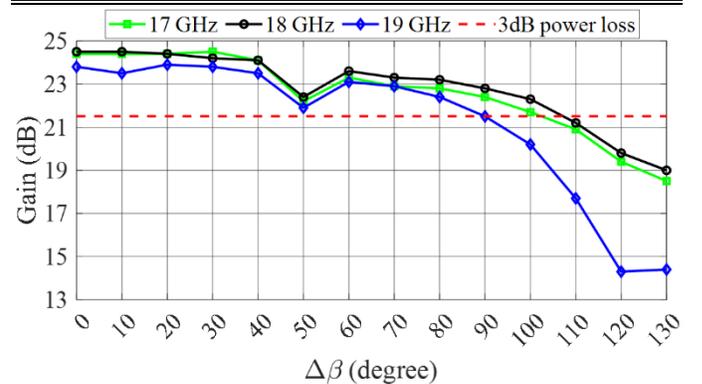
Fig. 21. The amount of gain loss as a function of the phase difference applied to the monopoles at frequencies 17, 18, and 19 GHz. The red dotted line indicates a 3 dB power loss at 18 GHz.

The pattern has rotated up to 47 degrees in the Azimuth plane, where the side lobe level is -11.1 dB and the half-power beamwidth is 9 degrees. Table. III compares the results for all cases of $\Delta\beta$ at the central frequency.

We analyze gain loss due to pattern rotation at 17, 18, and 19 GHz, as shown in Fig. 21. The φ-direction gain loss is significant, a trade-off for beam rotation. Fig. 21 shows a 3 dB gain loss at the center frequency (gain values: 21.4 dB at 17 GHz, 21.5 dB at 18 GHz) where the red line intersects the black curve. The phase difference $\Delta\beta$ enables azimuthal beam steering, as shown in Equation (3). For example, a $\Delta\beta$ of 108° achieves a 47° beam rotation with a 22.8 dB gain, per Table. III and Fig. 20b. A 3 dB gain loss allows steering up to ±47°, as indicated in Fig. 22, where the red dotted line and black curve intersect at the 3 dB threshold. Fig. 23 shows beam rotation at 18 GHz for different $\Delta\beta$ values in Cartesian coordinates.

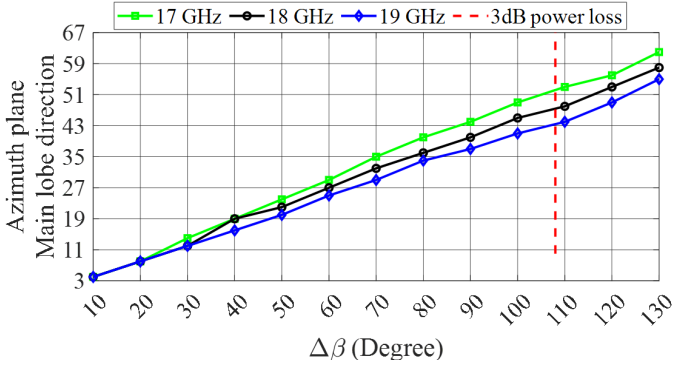

Fig. 22. shows the amount of beam rotation (in degrees) as a function of applying appropriate phase differences ($\Delta\beta$) to the monopoles for frequencies 17, 18, and 19 GHz. The red dotted line marks the 3 dB power loss at 18 GHz.

Note that by selecting the double size of antenna design, such as dimensions of 50 × 42 cm, we were able to rotate the beam to φ ≈ ±85° with minimal gain loss of approximately 0.6 dB. Therefore, if needed, by increasing the size of present structure (15×17 cm) in the y-direction, we can increase the angle of beam rotation in the φ-direction to significant values. Evidently, increasing the dimensions of antenna will increase the angle of rotation, but the antenna gain also decreases. Therefore, the trade-off among the gain, beam rotation, efficiency, and construction cost of antenna should be considered in detail.

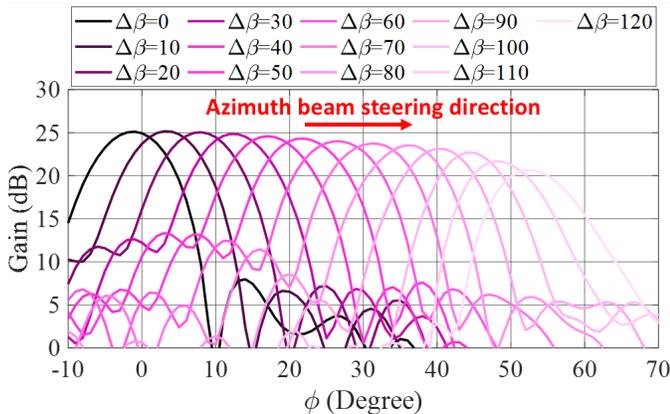

Fig. 23. Beam rotation in azimuth at 18 GHz for various phase shifts ($\Delta\beta$) applied to the monopoles.

### D. Monopole Spacing

The spacing between monopoles significantly influences the surface current distribution and, consequently, the antenna gain. To achieve optimal radiation enhancement, the monopole spacing should ideally be $\lambda/2$. To evaluate the sensitivity of the design to spacing variations, multiple spacing values were investigated. The results at the center frequency are presented in Fig. 24, where the horizontal axis denotes the monopole spacing, the left vertical axis indicates the antenna gain, and the right vertical axis shows the corresponding side lobe level.

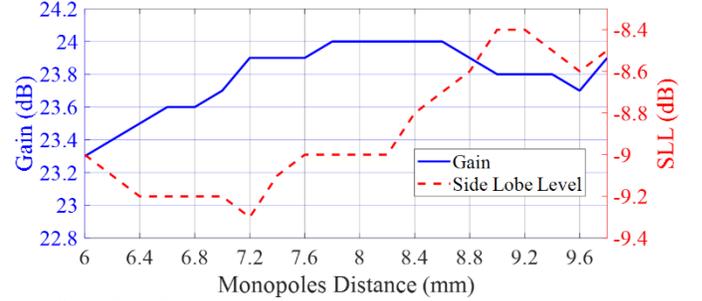

Fig. 24. Effect of monopole distance on surface current and gain at the central frequency where $\lambda_1/2 = 7.8$mm. The monopole distance, denoted as $d_M$, is shown in Fig. 1 and defined in Equation (4).

Fig. 24 shows that a monopole spacing of $\lambda/2$ maximizes gain and minimizes side lobe levels. Consequently, this spacing is used in the proposed design. The monopole distance, $d_M$, is depicted in Fig. 1 and defined in Equation (4).

### E. polarization

The antenna gain is plotted in Cartesian coordinates at the center frequency to analyze polarization.

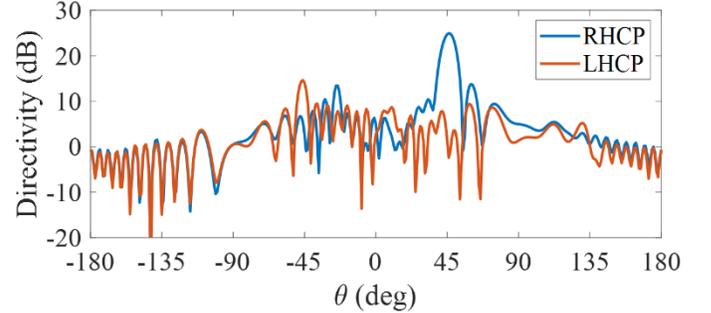

Fig. 25. Antenna gain in Cartesian coordinates on the φ=0° plane.

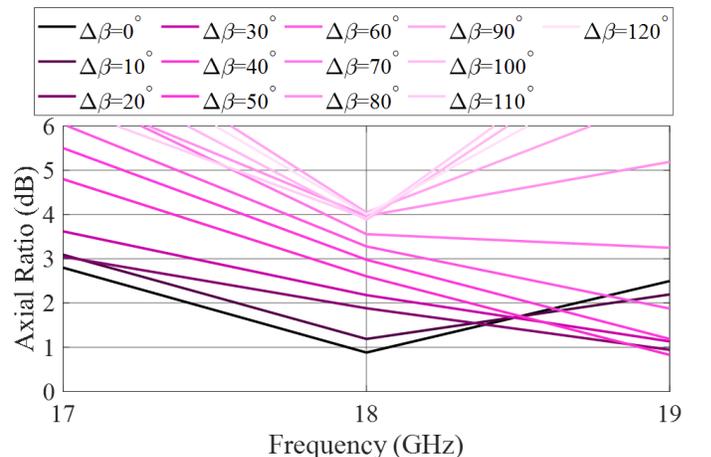

Fig. 26. Axial ratio in Cartesian coordinates on the φ=0° plane. The Axial Ratio for 18GHz is in the main lobe direction.

Fig. 25 displays the antenna pattern in the φ=0° plane, with a gain of 24.8 dB at θ=46°. It also confirms right-hand circular polarization (RHCP). Fig. 26 illustrates that the axial ratio in the main lobe direction for phase-shifted configurations remains consistently at or below 4 dB, indicating near-circular polarization during azimuthal beam steering, as calculated using Equation (27).

## V. BEAM FORMING NETWORK

The antennas in section III require a feeding network to operate at 17, 18, and 19 GHz. Initially, two Renesas F6212 8-channel beamforming ICs [36] were planned for steering beams in the φ-direction, but due to their high cost, a 1x16 SIW power divider was designed instead. This change eliminates φ-direction beam steering measurements but ensures minimal interference in the radiation pattern and produces a uniform wavefront. The network features one input and 16 in-phase output ports, with scattering parameters $S_{12}$ to $S_{117}$ around -12 dB. The design is shown in Fig. 27.

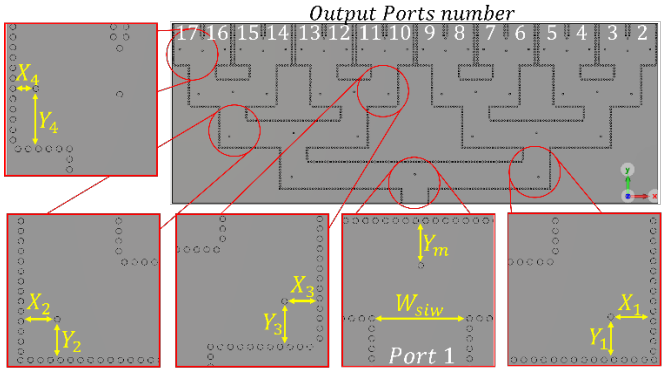

Fig. 27. Overall structure of the 1:16 SIW power divider. The distance of vias in the middle and corners of the structure and their positioning are $Y_m = 3.13mm$, $W_{siw} = 6.83mm$, $X_1 = 2.96mm$, $X_2 = 2.54mm$, $X_3 = 2.46mm$, $X_4 = 1.56mm$, $Y_1 = 3mm$, $Y_2 = 2.84mm$, $Y_3 = 3.21mm$, $Y_4 = 4.1mm$.

The structure measures approximately 12.3 × 4.5 cm on a Rogers RO4003C substrate. With via metallization, CST Studio Suite 2023 generates about 880,000 mesh elements. To halve simulation time, symmetry is exploited using a boundary condition ($H_t = 0$) on the Y-Z plane. Vias have a 0.4 mm diameter, with 0.7 mm center-to-center spacing.

As depicted in Fig. 27, the SIW feeding network has a width of $W_{siw} = 6.83mm$. The middle via, denoted $Y_m$, is fixed at 3.13mm across all layers. Corner vias, labeled $X_1$, $X_2$, $X_3$, $X_4$, $Y_1$, $Y_2$, $Y_3$, $Y_4$, vary by layer to enhance waveguiding, with indices indicating layer numbers. Port 1 serves as the input, while ports 2 to 17 are outputs.

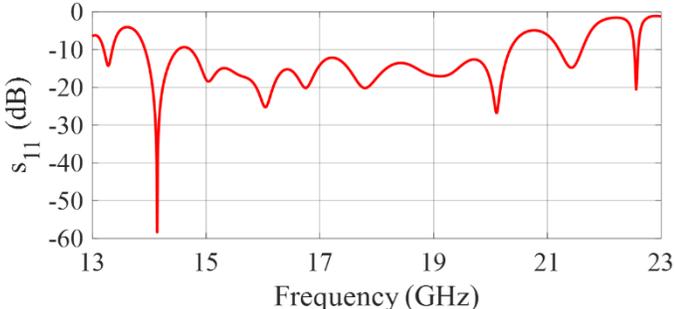

Fig. 28. $S_{11}$ of the 1:16 SIW power divider.

The feeding network, designed for a center frequency of 18 GHz, achieves a 14–20 GHz bandwidth (approximately 35.29%), with its $S_{11}$ shown in Fig. 28. Fig. 29 illustrates the electric field distribution at the center frequency. The scattering parameters (S-parameters) in Fig. 30a range from -12.9 dB to -13.6 dB, closely aligning with the theoretical target of -12 dB. Fig. 30b shows phase differences among output ports, confirming in-phase wave arrival with a maximum deviation of 3.57 degrees.

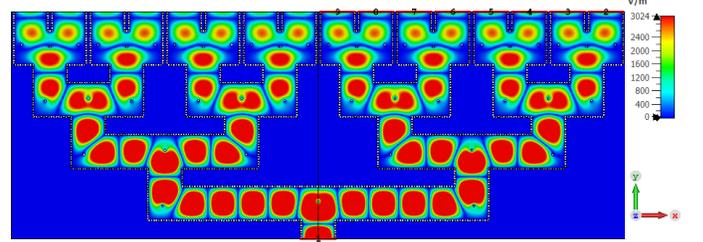

Fig. 29. Electric field distribution in the feeding network at the central frequency.

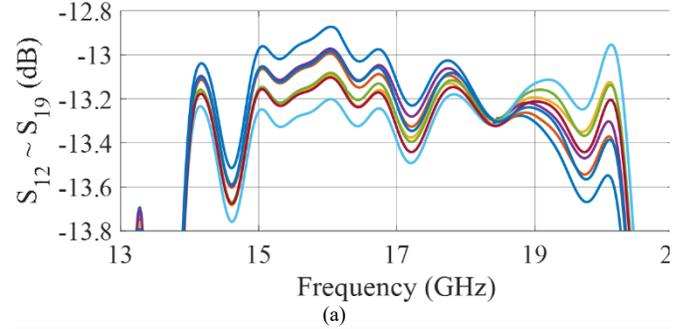

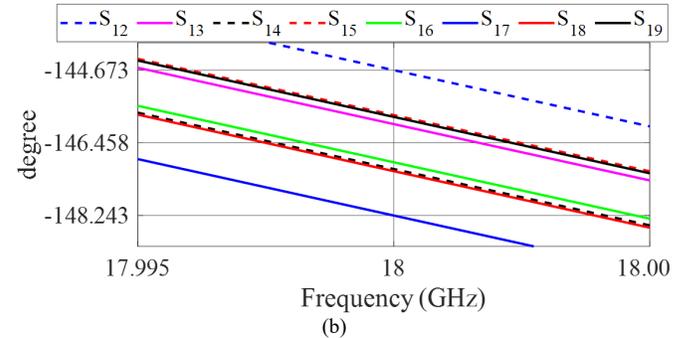

Fig. 30. (a) Values of $S_{12}$ to $S_{19}$ of the 1:16 SIW power divider. (b) Phase differences among the output ports.

## VI. FABRICATION OF ANTENNA SYSTEM AND MEASUREMENT

An efficient coupling device between the anisotropic antenna and the feeding network has been designed and fabricated. The overall structure, including top and back views, is illustrated in Fig. 31. Fig. 31b shows that the output of the feeding network is terminated by via metallization at the endpoint. The distance from the center of the coaxial cable to the center of the end vias, which served to close the output, is less than λ/4 (4.1 mm). Fig. 32 displays the fabricated antenna and power divider from top view.

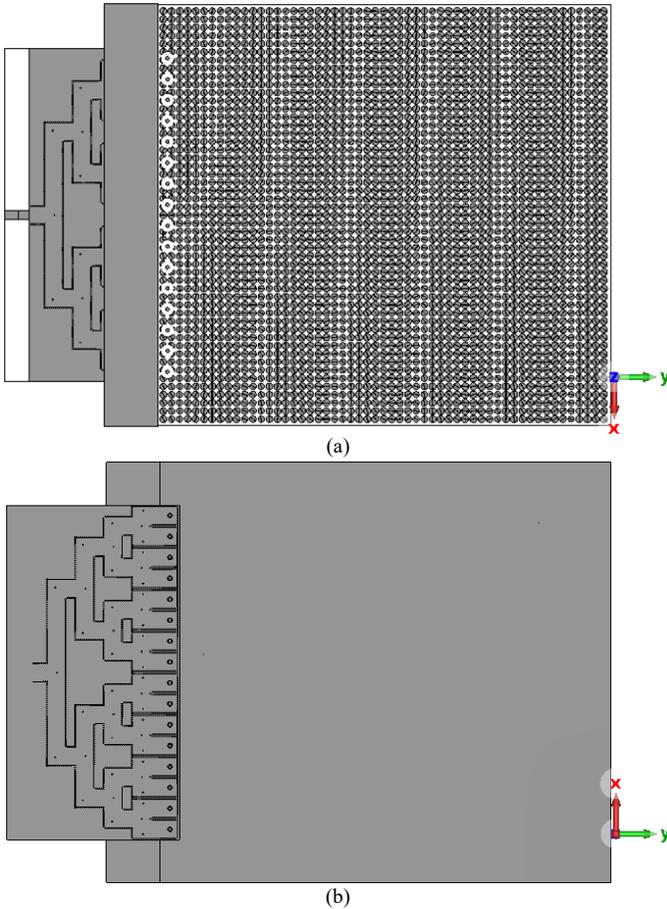

(a)

(b)

Fig. 31. Overall 2D structure view (a) top view (b) bottom view.

Coupling the antenna to the power divider may be implemented in three ways, as shown in Fig. 33:

1. Coaxial Cable.
2. SMA connector.
3. Slot.

Coaxial cables with connectors are expensive. A more cost-effective alternative is an SMA connector (part number 142-1000-001). For maximum savings, a slot can replace both the SMA connector and coaxial cable, providing the most economical solution for measurements.

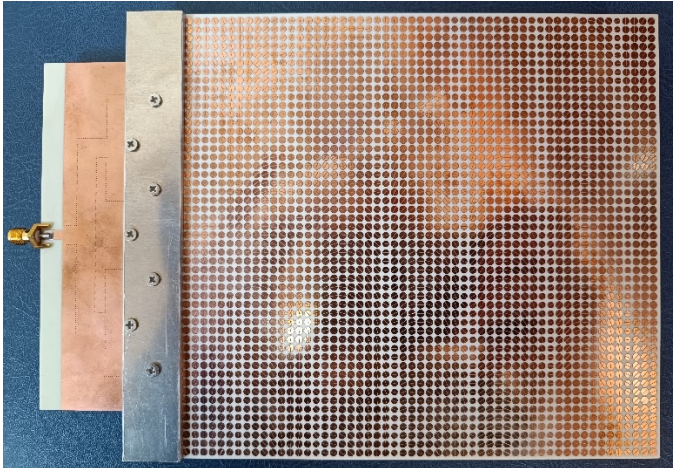

Fig. 32. Overall view of the fabricated antenna and 1:16 SIW power divider, shown from the top view.

We will compare the characteristics of these three options and summarize them in Table. VI.

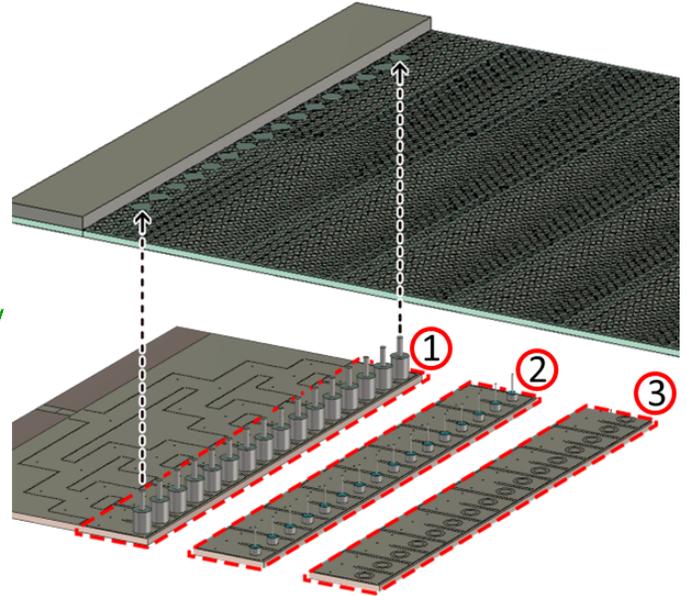

Fig. 33 Various methods for coupling an antenna with a 1:16 SIW power divider. 1- Coaxial cable; 2- SMA Connector; 3- Slot.

Fig. 34a shows the 3D radiation patterns of the slot-fed coupled antenna. Fig. 34b presents the 2D radiation patterns, comparing simulated and measured results in the θ-direction at 17, 18, and 19 GHz. Fig. 35 illustrates the simulated and measured $S_{11}$ and VSWR of the antenna, demonstrating strong agreement. The antenna achieves a 4.1 GHz bandwidth, spanning 15.6–19.7 GHz, as shown in Fig. 35.

Table. IV Comparison of the results for the different coupling methods between the antenna and the power divider: 1- Coaxial cable, 2- SMA connector, 3- Slot.

| | Bandwidth (GHz) | Gain (dB) | SLL In E-plane (dB) | SLL in H-plane (dB) | HPBW In E-plane (deg.) | HPBW In H-plane (deg.) | Rad. Eff. |
|---|---|---|---|---|---|---|---|
| 1 | 15.5~20.2 | 21.5 | -10.4 | -6.7 | 10.6 | 9.9 | %71 |
| 2 | 16~20.4 | 24.5 | -11 | -12.7 | 9.5 | 9.4 | %74 |
| 3 | 16.1~20.5 | 21.9 | -7 | -11.4 | 8.6 | 9.1 | %72 |

The antenna beam angles are 42°, 48°, and 55°, with corresponding beamwidths of 8.9°, 9.8°, and 12.3°, and side lobe levels of -5 dB, -6 dB, and -5.8 dB, respectively. Using monopoles instead of slots for feeding could further reduce side lobe levels. In conclusion, Table. V presents a comparison of this work with similar recent studies. A summary of the overall characteristics of the coupled antenna with the power divider for each of the three options at the center frequency is compared in Table. IV.

Table. V Comparison with other similar studies in the recent literature.

| Paper | Gain (dB) | Frequency (GHz) | Structure dimension (cm) | Beam Steering Coverage in Elevation by | | Beam Steering Coverage in Azimuth by | | Rad. Efficiency | Polarization | Implemented methode |
|---|---|---|---|---|---|---|---|---|---|---|
| | | | | Overal steering | 3 dB Power Loss | Overal steering | 3 dB Power Loss | | | |
| [37] | 10 | 10 | 12×12 | 0°~75° | 75° | ±60° | ≈ ±40° | N/A | N/A | Waveguide-fed metasurface antenna |
| [18] | 18 | 90 | 10×10 | 111°~128° | N/A | ±18° | N/A | N/A | N/A | The theory of holographic |
| [38] | 22.3 | 27.75~29.6 | 28×28 | 0° ~ 66° | 62° | ±180° | ≈ ±15° | N/A | Circular | The Risley prism beam-steering concept using only a single flat prism |
| [39] | 31.9 | 94 | 5×5 | 0° ~ 45° | N/A | 10° ~ 15° | N/A | N/A | N/A | Holographic Modulation + Integrated Horns |
| [40] | 14.2 | 24, 38 | 11.8×11.8 | 0° ~ 40° | N/A | - | N/A | %79 ~ %86 | N/A | Selecting a specific port of MIMO antenna |
| [41] | 6.25 | 28.8~31.8 | 15.4×15.4 | N/A | N/A | N/A | N/A | %68 ~ %90 | N/A | Multilayer structure of liquid crystal (LC) material |
| This Work | 25.1 | 17~19 | 15×18 | 30° ~ 65° | 30° ~ 65° | ±58° | ±42° | %86 | Circular | The holographic principle combined with array theory |

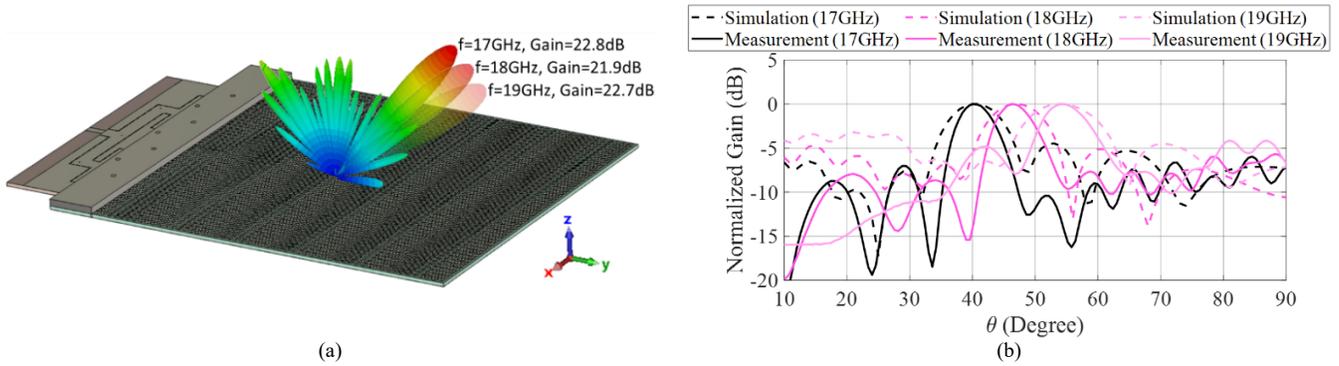

(a)                                                                 (b)

Fig. 34. (a) 3D Radiation patterns simulation of the overall antenna structure in the θ-direction at 17 GHz, 18 GHz, and 19 GHz. (b) 2D radiation pattern simulation in the θ-direction at 17 GHz, 18 GHz, and 19 GHz validated by measurements.

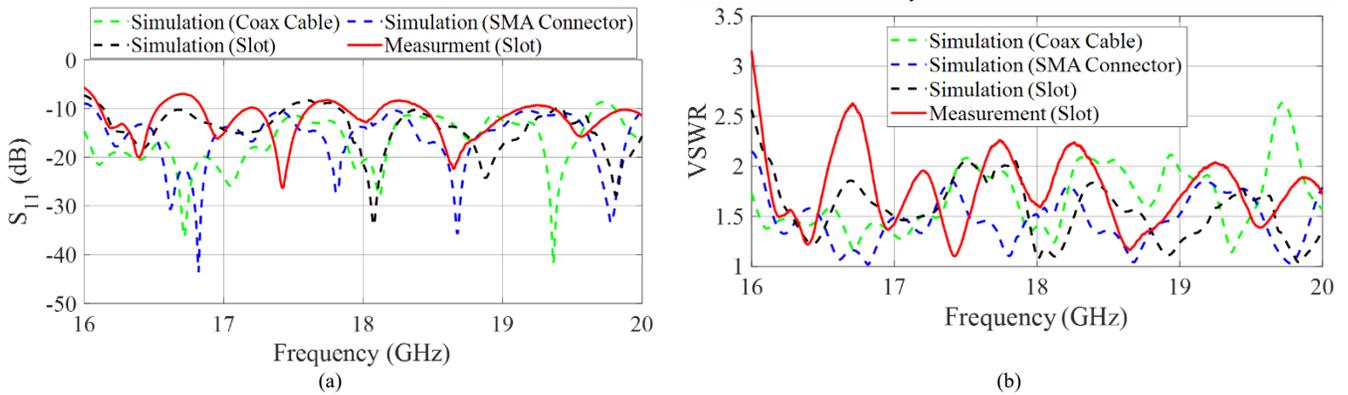

(a)                                                                 (b)

Fig. 35. (a) $S_{11}$ and (b) VSWR of the overall antenna structure for the simulated coaxial cable, SMA connector, and slot cases, compared to the measured results for the slot case.

Table. VI Summary of the three options mentioned for coupling power between the antenna and power divider. The dimensions of the slot are $R_{in} = 1.3 mm, and R_{out} = 2.1 mm$.

| | Coupling interface | Total cost | Simulated | Manufactured | Figure |
|---|---|---|---|---|---|
| 1 | Coaxial Cable | High | Yes | No | 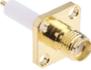 |
| 2 | SMA Connector | Medium | Yes | No | 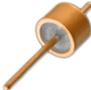 |
| 3 | Slot | Low | Yes | Yes | 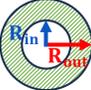 |

## VII. CONCLUSION

This paper introduces a 15 × 17 cm² holographic antenna for $K_u$ band operation at 18 GHz. Using phase-shifted monopoles, it achieves ±58° beam steering in the φ-direction and 35° to 57° in the θ-direction via 16–20 GHz frequency tuning. The antenna offers 24.5 dB gain, 10° HPBW, 22.22% bandwidth, and circular polarization. Beam broadening factors are 1.1 (azimuth) and 1.4 (elevation), with side lobe levels at -10 dB and -12.9 dB. A 1:16 SIW power divider feeds the 16-monopole array. Validated by CST 2023 simulations and measurements, with MATLAB-CST integration optimizing unit cell arrangement, this fabricated design shows strong potential for 5G and 6G applications, advancing holographic antenna technology.

## VIII. ACKNOWLEDGMENT

## IX. REFERENCES

[1] F. Yang, P. Pitchappa, and N. Wang, "Terahertz reconfigurable intelligent surfaces (RISs) for 6G communication links," *Micromachines,* vol. 13, no. 2, p. 285, 2022.

[2] G. Minatti, F. Caminita, M. Casaletti, and S. Maci, "Spiral leaky-wave antennas based on modulated surface impedance," *IEEE Transactions on antennas and propagation,* vol. 59, no. 12, pp. 4436-4444, 2011.

[3] S. Maci, "Design of modulated metasurface antennas," *2017 IEEE International Symposium on Antennas and Propagation & USNC/URSI National Radio Science Meeting,* pp. 1939-1940, 2017.

[4] M. Faenzi *et al.*, "Metasurface antennas: New models, applications and realizations," *Scientific reports,* vol. 9, no. 1, p. 10178, 2019.

[5] H. Oraizi, A. Amini, A. Abdolali, and A. M. Karimimehr, "Design of wideband leaky-wave antenna using sinusoidally modulated impedance surface based on the holography theory," *IEEE Antennas and Wireless Propagation Letters,* vol. 17, no. 10, pp. 1807-1811, 2018.

[6] M. M. Moeini, H. Oraizi, and A. Amini, "Design of wide-band 2-d holographic antenna using wedge reflector," in *2018 48th European Microwave Conference (EuMC)*, 2018: IEEE, pp. 428-431.

[7] M. M. Moeini, H. Oraizi, and A. Amini, "Collimating cylindrical surface leaky waves for highly improved radiation characteristics of holograms," *Physical Review Applied,* vol. 11, no. 4, p. 044006, 2019.

[8] A. Amini, H. Oraizi, M. Movahediqomi, and M. M. Moeini, "Adiabatic Floquet-Wave Solutions of Temporally Modulated Anisotropic Leaky-Wave Holograms," *arXiv preprint arXiv:2206.00797,* 2022.

[9] A. Amini and H. Oraizi, "Temporally modulated one-dimensional leaky-wave holograms," *Scientific Reports,* vol. 12, no. 1, p. 8488, 2022.

[10] A. Gupta and R. K. Jha, "A survey of 5G network: Architecture and emerging technologies," *IEEE access,* vol. 3, pp. 1206-1232, 2015.

[11] A. Sufyan, K. B. Khan, O. A. Khashan, T. Mir, and U. Mir, "From 5G to beyond 5G: A comprehensive survey of wireless network evolution, challenges, and promising technologies," *Electronics,* vol. 12, no. 10, p. 2200, 2023.

[12] H. Ma, J.-S. Kim, J.-H. Choe, and Q.-H. Park, "Deep-learning-assisted reconfigurable metasurface antenna for real-time holographic beam steering," *Nanophotonics,* vol. 12, no. 13, pp. 2415-2423, 2023.

[13] T.-H. Lin and R.-B. Hwang, "A Beam-steerable Holographic Antenna," in *2020 International Workshop on Electromagnetics: Applications and Student Innovation Competition (iWEM)*, 2020: IEEE, pp. 1-2.

[14] C. Huang *et al.*, "Holographic MIMO surfaces for 6G wireless networks: Opportunities, challenges, and trends," *IEEE Wireless Communications,* vol. 27, no. 5, pp. 118-125, 2020.

[15] C. Rusch, "Holographic Antennas," in *Handbook of Antenna Technologies*, Z. N. Chen, D. Liu, H. Nakano, X. Qing, and T. Zwick Eds. Singapore: Springer Singapore, 2016, pp. 2689-2725.

[16] C. Rusch, S. Beer, P. Pahl, H. Gulan, and T. Zwick, "Electronic beam scanning in two dimensions with holographic phased array antenna," in *2013 International Workshop on Antenna Technology (iWAT)*, 2013: IEEE, pp. 23-26.

[17] A. Amini, H. Oraizi, M. Hamedani, and A. Keivaan, "Wide-band polarization control of leaky waves on anisotropic holograms," *Physical Review Applied,* vol. 13, no. 1, p. 014038, 2020.

[18] S. Beer, P. Pahl, T. Zwick, and S. Koch, "Two-dimensional beam steering based on the principle of holographic antennas," in *2011 International Workshop on Antenna Technology (iWAT)*, 2011: IEEE, pp. 210-213.

[19] M. C. Johnson and B. Rothaar, "Beam shaping for reconfigurable holographic antennas," ed: Google Patents, 2017.

[20] M. A. Jamlos, M. Khairi, S. N. Ariffah, W. A. Mustafa, S. Z. S. Idrus, and A. Muhammad, "5.8 GHz Circular Polarized Microstrip Feeding Antenna for Solar Panel Application," in *IOP Conference Series: Materials Science and Engineering*, 2020, vol. 932, no. 1: IOP Publishing, p. 012077.

[21] Y. Jia *et al.*, "In-band RCS reduction of circularly polarized microstrip antenna array based on a novel feed network," *IEEE Transactions on Antennas and Propagation,* 2024.

[22] A. Elshafiy, K. Rose, and A. Sampath, "On optimal beam steering directions in millimeter wave systems," in *ICASSP 2019-2019 IEEE International Conference on Acoustics, Speech and Signal Processing (ICASSP)*, 2019: IEEE, pp. 4709-4713.

[23] K. Honda, T. Fukushima, and K. Ogawa, "Full-azimuth beam steering MIMO antenna arranged in a daisy chain array structure," *Micromachines,* vol. 11, no. 9, p. 871, 2020.

[24] "https://www.mathworks.com/products/matlab.html." (accessed.

[25] "https://www.comsol.com/product-download." (accessed.

[26] "CST Studio Suite, Computer Simulation Technolog ag, https://www.cst.com/." (accessed.

[27] C. A. Balanis, *Antenna theory: analysis and design*. John wiley & sons, 2016.

[28] M. M. Moeini, H. Oraizi, A. Amini, and V. Nayyeri, "Wide-band beam-scanning by surface wave confinement on leaky wave holograms," *Scientific reports,* vol. 9, no. 1, p. 13227, 2019.

[29] S. Tretyakov, *Analytical modeling in applied electromagnetics*. Artech House, 2003.

[30] G. Minatti, F. Caminita, E. Martini, and S. Maci, "Flat optics for leaky-waves on modulated metasurfaces: Adiabatic Floquet-wave

analysis," *IEEE Transactions on Antennas and Propagation,* vol. 64, no. 9, pp. 3896-3906, 2016.

[31] M. Teniou, H. Roussel, N. Capet, G.-P. Piau, and M. Casaletti, "Implementation of radiating aperture field distribution using tensorial metasurfaces," *IEEE Transactions on Antennas and Propagation,* vol. 65, no. 11, pp. 5895-5907, 2017.

[32] G. Minatti *et al.*, "Modulated metasurface antennas for space: Synthesis, analysis and realizations," *IEEE Transactions on Antennas and Propagation,* vol. 63, no. 4, pp. 1288-1300, 2014.

[33] A. Amini and H. Oraizi, "Adiabatic floquet-wave expansion for the analysis of leaky-wave holograms generating polarized vortex beams," *Physical Review Applied,* vol. 15, no. 3, p. 034049, 2021.

[34] D. H. Werner and D.-H. Kwon, *Transformation electromagnetics and metamaterials.* Springer, 2013.

[35] V. G. Ataloglou, S. Taravati, and G. V. Eleftheriades, "Metasurfaces: physics and applications in wireless communications," *National Science Review,* vol. 10, no. 8, p. nwad164, 2023.

[36] Renesas. "Beamforming ICs for SATCOM and radar active antennas." https://www.renesas.com/en/document/prb/beamforming-ics-satcom-and-radar-active-antennas?r=1164871 (accessed January 7, 2025.

[37] M. Boyarsky, T. Sleasman, M. F. Imani, J. N. Gollub, and D. R. Smith, "Electronically steered metasurface antenna," *Scientific reports,* vol. 11, no. 1, p. 4693, 2021.

[38] Z. Zhang, Y. C. Zhong, H. Luyen, J. H. Booske, and N. Behdad, "A low-profile, Risley-prism-based, beam-steerable antenna employing a single flat prism," *IEEE Transactions on Antennas and Propagation,* vol. 70, no. 8, pp. 6646-6658, 2022.

[39] O. Yurduseven *et al.*, "Multibeam Si/GaAs holographic metasurface antenna at W-band," *IEEE Transactions on Antennas and Propagation,* vol. 69, no. 6, pp. 3523-3528, 2020.

[40] B. T. Malik, S. Khan, and S. Koziel, "Beam steerable MIMO antenna based on conformal passive reflective metasurface for 5G millimeter wave applications," *Scientific Reports,* vol. 14, no. 1, p. 24086, 2024.

[41] P. Chen, D. Wang, L. Wang, L. Liu, and Z. Gan, "Liquid crystal-based reconfigurable antenna for 5G millimeter-wave," *Scientific Reports,* vol. 14, no. 1, p. 16646, 2024.